# $MgB_2$ and $Mg_{1-x}Al_xB_2$ single crystals: high-pressure growth and physical properties

J.Karpinski[a], S.M.Kazakov[a], J.Jun[a], N.D.Zhigadlo[a], M.Angst[a], R.Puzniak[b], A.Wisniewski[b]

[a]*Solid State Physics Laboratory ETH 8093-Zurich, Switzerland,*
[b]*Institute of Physics, Polish Academy of Sciences, 02-668 Warsaw, Poland,*

**Abstract.**

Single crystals of $MgB_2$ have been grown with a high-pressure cubic anvil technique. They grow via the peritectic decomposition of the $MgNB_9$ ternary nitride. The crystals are of a size up to $2 \times 1 \times 0.1\ mm^3$ with a weight up to 230μg. Typically they have transition temperatures between 38 and 38.6 K with a width of 0.3-0.5 K. Investigations of the *P-T* phase diagram prove that the $MgB_2$ phase is stable at least up to 2190 °C at high hydrostatic pressure in the presence of Mg vapor under high pressure. Substitution of aluminum for magnesium in single crystals leads to stepwise decrease of $T_c$. This indicates a possible appearance of superstructures or phases with different $T_c$'s. The upper critical field decreases with Al doping.

# 1. Introduction

MgB$_2$ has anisotropic two-band electronic structure, leading to the presence of two distinct energy gaps [1,2]. The two gaps in MgB$_2$ arise from different strengths of the electron-phonon coupling in the σ and π bands. This causes unconventional properties, like the high critical temperature of 39 K and a temperature and field dependent anisotropy, which does not fit to the Ginzburg-Landau model with a constant anisotropy [3]. Substitutions for Mg or B can influence superconducting parameters. Anisotropic properties have to be studied on single crystals. Conventional methods of crystal growth, like a growth from high temperature solutions in metals (Al, Mg, Cu, etc.) at ambient pressure, used for other borides, did not work for MgB$_2$. Since MgB$_2$ melts non-congruently, it is also not possible to grow crystals from a stoichiometric melt. MgB$_2$ crystals can be grown from a solution in Mg at high pressure. However, solubility of MgB$_2$ in Mg is very low, therefore crystals grown using this method are rather small, well below 1x1mm$^2$.

# 2. Crystal growth

The goal of our experiments was to grow pure and substituted MgB$_2$ single crystals for investigations of intrinsic superconducting parameters. We have applied high pressure, high temperature crystal growth method developed recently in our laboratory [4,5], using decomposition reaction of the MgNB$_9$ nitride. Typically the BN crucible with a mixture of Mg, B and BN has been placed in the pyrophyllite cube containing graphite heater in the cubic anvil device. The conditions of crystal growth were: pressure between 20 and 30 kbars, maximum temperature between 1800 and 2190 $^o$C. Temperature profiles include heating up to maximum temperature during 1-2 h, dwelling for 0.1-12 h and cooling during 1-3 h.  Single crystals of MgB$_2$ and BN were grown from the reactions in the ternary Mg-N-B system. This process is a combination of several reactions, which lead to growth of MgB$_2$ crystals of a size up to 2x1x0.1mm$^3$ (Fig.1). Crystal growth was performed at temperatures up to 2190 $^o$C, which proved the stability of the MgB$_2$ at such high temperature.

# 3. Substitution of Al for Mg

Substitutions in MgB$_2$ have been studied up to now only in polycrystalline samples. The most investigated are Al substitution for Mg and C substitution for B. Both of them lead to decrease of $T_c$. From the fundamental point of view it is very interesting to know how these substitutions influence basic superconducting parameters such as upper critical fields, anisotropy, both energy gaps, band structure etc. These investigations are best performed on single crystals. In this paper we present preliminary results of the influence of Al substitutions on $T_c$. For the substitution we have replaced in the precursor part of Mg with Al. The rate of Al replacement for Mg in the precursor varies from 5 to 50% in various experiments. The process of crystal growth has been performed in the same way as for the non-substituted crystals. The amount of Al in single crystals has been investigated with EDX (Energy Dispersive X-ray) analysis. Aluminium content in $Mg_{1-x}Al_xB_2$ crystals is significantly lower than this in the precursor. Investigated crystals contain up to x = 0.28 Al substituted for Mg. The results are shown in Fig. 2. Measurements of $T_c$ (effective) show an unexpected result: $T_c$'s values do not vary in a continuos way with a variation of Al content. $T_c$'s seem to have discrete values, decreasing stepwise for various samples. Some of crystals have the same $T_c$ for various Al content. For the same Al content there are crystals with different $T_c$'s. The reason of the stepwise change of $T_c$ may be the appearance of various ordered phases. However, HRTEM investigations performed on crystals with x = 0.02-0.03 of Al did not show ordering [6]. These results are different from the ones obtained on polycrystalline samples [7], which show continuous change of $T_c$ on Al content. The possible reason of this discrepancy can be that polycrystalline samples contain a collection of many crystallites and investigations provide average values for many crystallites only. Investigations of lattice constants show continuous decrease of $a$ from 3.0844(4) to 3.0677(4) Å and $c$ from 3.5195(7) to 3.426(2) Å corresponding to Al content increase from x = 0 to 0.28 and decrease of $T_c$ from 38.5 to 14.9 K. Magnetic investigations with a torque magnetometer show a decrease of the upper critical field with Al substitution.

Figure captions.

Fig.1. Single crystals of $MgB_2$. Scale size is 1mm.

Fig.2. $T_c$ versus Al content (EDX) in $Mg_{1-x}Al_xB_2$ crystals grown at T = 1860 $^o$C, compared with $T_c$ dependence for polycrystalline samples [7].

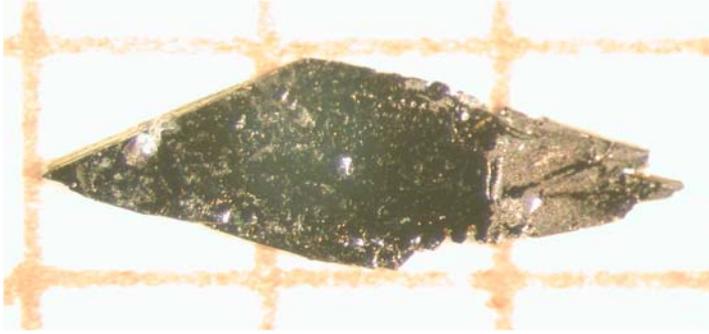
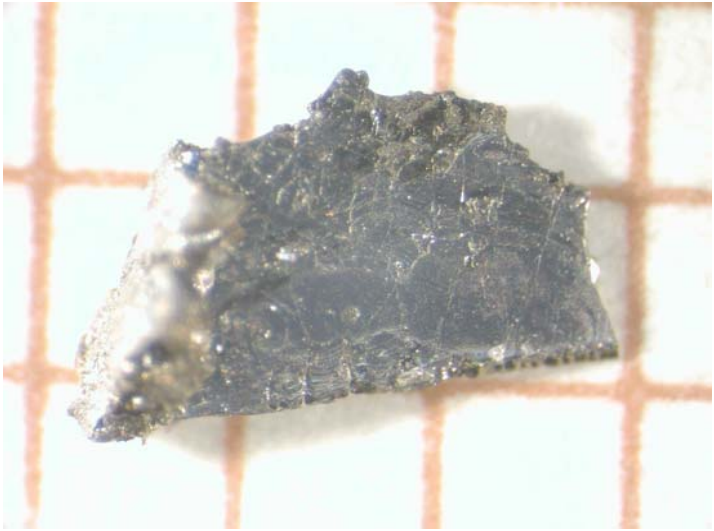

Figure 1.

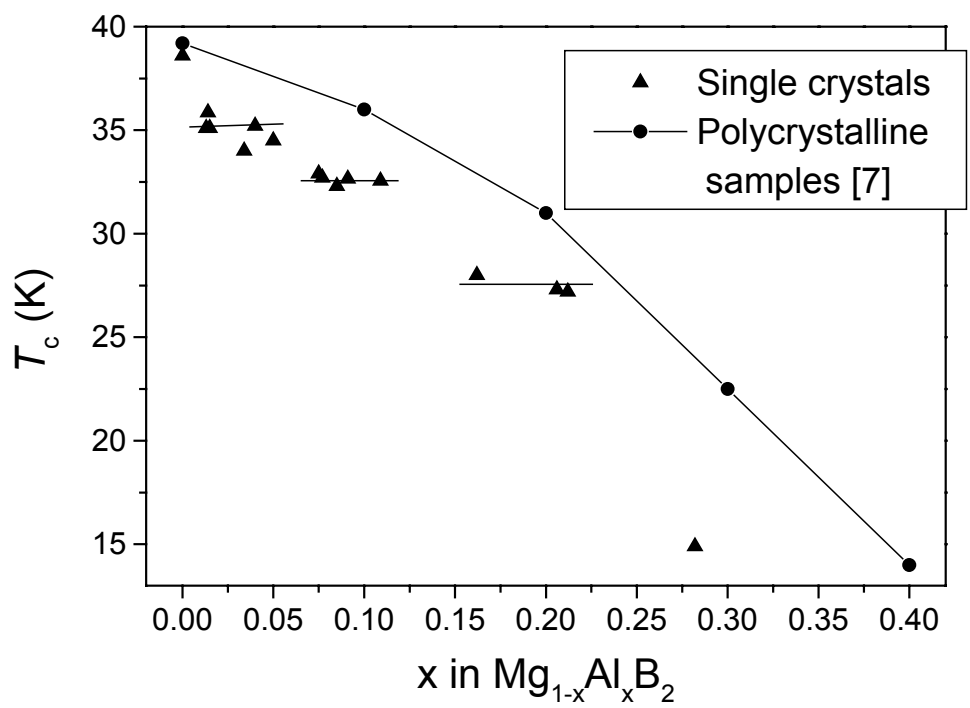

Figure 2.